\documentclass[aps,nofootinbib,showpacs]{revtex4}
\usepackage{graphicx}

\begin{document}

\preprint{DTP-MSU/01-23 \\ hep-th/0110164}

\title{Supergravity Fluxbranes in Various Dimensions}

\author{Chiang-Mei Chen}
 \email{cmchen@phys.ntu.edu.tw}
 \affiliation{Department of Physics, National Taiwan University,
Taipei 106, Taiwan, R.O.C.}

\author{Dmitri V. Gal'tsov}
 \email{galtsov@grg.phys.msu.su}
 \affiliation{Department of Theoretical Physics, Moscow State University,
119899, Moscow, Russia}

\author{Paul M. Saffin}
 \email{p.m.saffin@durham.ac.uk}
 \affiliation{Centre of Particle Theory, Department of
Mathematical Sciences, University of Durham, South Road, Durham
DH1 3LE, United Kingdom}

\date{\today}

\begin{abstract}
We investigate fluxbrane solutions to the Einstein-antisymmetric
form-dilaton theory in arbitrary space-time dimensions for a
transverse space of cylindrical topology $S^k\times R^n$,
corresponding to smeared and unsmeared solutions. A master
equation for a single metric function is derived. This is a
non-linear second-order ordinary differential equation admitting
an analytic solution, singular at the origin, which serves as an
attractor for globally regular solutions, whose existence is
demonstrated numerically. For all fluxbranes of different levels
of smearing the metric function diverges at infinity as the same
power of the radial coordinate except for the maximally smeared
case, where a global solution is known in closed form and can be
obtained algebraically using U-duality. The particular cases of F6
and F3 fluxbranes in $D=11$ supergravity and fluxbranes in IIA,
IIB supergravities are discussed.
\end{abstract}


\maketitle

\section{Introduction}
Recently there has been an upsurge of interest in fluxbrane
solutions of supergravity theories which generalize the well-known
Melvin magnetic universe \cite{Me64}. The Melvin solution is
supported by a one-form potential and is currently denoted as F1,
$D=4$; its generalization to higher dimensions, including a
dilaton, was performed by Gibbons and Maeda {\cite{GiMa88}. The F1
fluxbrane was later shown to have a nice interpretation as a
modding of flat space in one dimension higher
\cite{DoGaKaTr94,DoGaGiHo95,DoGaGiHo96,CoGu00}. Another generating
technique which works for Fp-branes supported by higher rank
antisymmetric potentials was suggested in \cite{GaRy98}, it
consists of using the Harrison transformation from a U-duality
group arising in the dimensional reduction of the
Einstein-antisymmetric form-dilaton action with respect to a
spacelike Killing direction. A more sophisticated
gravity/supergravity duality based on the reinterpretation of the
duals of the Kaluza-Klein two-forms as higher rank antisymmetric
forms of certain supergravities was used to construct intersecting
fluxbranes \cite{ChGaSh99b,ChGaSh99a} of M-theory. Later on direct
methods of solving the Einstein equations were applied to
investigate the fluxbrane solutions \cite{Sa01a,GuSt01}, see also
other related work
\cite{CoHeCo01,Em01,Sa01b,brecher01,Mo01,Su01,Ur01}. Various
physical applications for fluxbranes in string and M-theory have
been discussed: construction of exact string backgrounds
\cite{RuTs01,Ts01}, duality between type 0A and IIA string
theories \cite{CoGu00}, gravitating version of the dielectric
effect \cite{CoHeCo01,Em01,brecher01}.

Here we concentrate on the topology of the transverse space in
fluxbrane solutions. In the existing literature (with an exception
of \cite{GaRy98,ChGaSh99a}) two different transverse space
structures were considered. The first was the product of the
radial line by a circle with the remaining transverse directions
being flat, corresponding to a smearing of the F1 (Melvin)
solution. The other was the case of the fully spherically
symmetric transverse space. We will show that a more general class
of fluxbrane solutions exists interpolating between these two
extremes, namely solutions with the transverse space being the
product of $k$-dimensional sphere with the remaining dimensions
being flat, leading to different levels of smearing. Therefore,
there exist F6 fluxbranes with $k=1,2,3$ and F3 fluxbranes with
$k=1,2,3,4,5,6$ in M-theory (this is very similar to the case of
the ordinary M2 and M5 branes), as well as various smeared
fluxbranes in type IIA and IIB theories. All of them are governed
by a single second order master equation which we investigate both
analytically and numerically.

\section{General F$\textnormal{p}$-branes}
In this section we analyse equations governing Fp-branes
associated with the flux of a $q$-form field strength. The system
contains a graviton, a $q$-form field strength, F$_{[q]}$, and a
dilaton scalar, $\phi$, coupled to the form field with the
coupling constant $a$. This is a general framework which
encompasses the bosonic sector of various supergravity theories,
coming from a truncation of the low energy limit of M-theory and
string theories, by a certain choice of the dimension $D$, the
rank of form field $q$, and the dilaton coupling $a$. In the
Einstein frame, the action is given by
\begin{equation}\label{action}
S = \int d^D x \sqrt{-g} \left( R - \frac12 \partial_\mu \phi
\partial^\mu \phi - \frac1{2\, q!} \, {\rm e}^{a\phi} \, F_{[q]}^2
\right).
\end{equation}
This action is invariant under the following discrete S-duality:
\begin{equation} \label{duality}
g_{\mu\nu} \to g_{\mu\nu}, \qquad F \to {\rm e}^{-a\phi} \ast F,
\qquad \phi \to -\phi,
\end{equation}
where $\ast$ denotes a $D$-dimensional Hodge dual. This may be
used to construct electric versions of magnetic fluxbranes and
vice versa. The equations of motion, derived from the variation
of the action with respect to the individual fields, are
\begin{eqnarray}
R_{\mu\nu} - \frac12 \partial_\mu \phi \partial_\nu \phi -
\frac{{\rm e}^{a\phi}}{2(q-1)!} \left[
F_{\mu\alpha_1\cdots\alpha_{q-1}}
F_\nu{}^{\alpha_1\cdots\alpha_{q-1}}- \frac{q-1}{q(D-2)}
F_{[q]}^2 \, g_{\mu\nu} \right] &=& 0, \\ \label{Ein}
\partial_\mu \left( \sqrt{-g} \, {\rm e}^{a\phi} \,
F^{\mu\nu_1\cdots\nu_{q-1}} \right) &=& 0, \\ \label{form}
\frac1{\sqrt{-g}}\, \partial_\mu \left( \sqrt{-g} \partial^\mu
\phi \right) - \frac{a}{2\, q!} {\rm e}^{a\phi} F_{[q]}^2 &=&
0.\label{dil}
\end{eqnarray}

We study fluxbranes with world volumes possessing the $p+1$
dimensional Poincar\'e invariance and with a transverse space
having the SO($k$), $k\leq q-1$, rotational symmetry, the
remaining $q-k-1$ dimensions being (conformally) flat. Obviously,
in $D$ dimensions, $p=D-q-1$. For $k= q-1$ the transverse space
is spherically symmetric, while for lower $k$ one deals with
cylindrical symmetry (smeared fluxbranes). With this in mind we
choose the metric
\begin{equation}\label{metric}
ds^2 = {\rm e}^{2A} (-dt^2 + dx_1^2 + \cdots + dx_p^2) + {\rm
e}^{2B} dr^2 + {\rm e}^{2C} (r^2 d\Omega_k^2 + dy_1^2 + \cdots +
dy_{q-k-1}^2),
\end{equation}
parameterized by three $r$-dependent functions $A(r), B(r)$ and
$C(r)$. An alternative gauge $B=0$ can be used in the
spherically-symmetric case $k=q-1$, but we prefer the curvature
coordinates for the cylindrical transverse space, in which case
$B\neq 0$.

With this ansatz, the equation for the form field (\ref{form}),
can easily be solved giving
\begin{equation}\label{SolF}
F_{[q]} = 2 b  \, {\rm e}^{H-2(p+1)A+2B-a\phi}\, dr \wedge
\epsilon_{[k]} \wedge dy_1 \wedge \cdots \wedge dy_{q-k-1},
\end{equation}
where $b$ is the field strength parameter, $\epsilon_{[k]}$
denotes the unit volume form of the round sphere $S^k$, and
\begin{equation}\label{D}
H = (p+1)A - B + (q-1) C + k \ln r.
\end{equation}
Substituting this into the Eq.(\ref{dil}) one obtains the
following dilaton equation
\begin{equation}\label{EqPhi}
\phi'' + H'\phi' - 2 a b^2 \, {\rm e}^{-2(p+1)A+2B-a\phi} = 0,
\end{equation}
where primes denote derivatives with respect to $r$.

To derive the equations for the metric functions $A,\,B,\,C$ one
calculates first the Ricci tensor for the metric (\ref{metric}),
the non-vanishing components being
\begin{eqnarray}
- R_{tt} &=& R_{xx} = - {\rm e}^{2A-2B} ( A'' +  H'A'), \\
- R_{rr} &=& (p+1)(A'' + A'^2 - A'B') + (q-1) (C'' + C'^2 - B'C')
+ k r^{-1}(2C'-B'), \label{Rrr} \\
R_{yy} &=& - {\rm e}^{2C-2B} ( C'' + H'C'), \label{Ryy} \\
R_{\omega\omega} &=&  \left[ r^2 R_{yy} + k - 1 - {\rm e}^{2C-2B}
(r H'-1) \right]\, \bar g_{\omega\omega}. \label{Romegaomega}
\end{eqnarray}
Here $\bar g_{\omega\omega}$ denotes the metric of the sphere
$S^k$, $d\Omega_k^2 = \bar g_{\omega\omega} d\omega d\omega$. In
the case of the full spherical symmetry of the transverse space
$k=q-1$ there is no $R_{yy}$ component of the Ricci tensor, while
the $R_{yy}$ term in (\ref{Romegaomega}) has to be understood as
a shorthand for the expression on the right hand side of
(\ref{Ryy}). Finally, the Einstein equations (\ref{Ein}) reduce
to the following set of equations
\begin{eqnarray}
A'' +  H'A' - \frac{2(q-1)b^2}{D-2} {\rm e}^{-2(p+1)A+2B-a\phi}
&=& 0, \label{EqA} \\
C'' +  H'C' + \frac{2 p b^2}{D-2} {\rm e}^{-2(p+1)A+2B-a\phi} &=&
0, \label{EqC} \\
(p+1)(A'' + A'^2 - A'B') + (q-1) (C'' + C'^2 - B'C') + k
r^{-1}(2C'-B') && \nonumber \\
+ \frac12 \phi'^2 + \frac{2 p b^2}{D-2} {\rm
e}^{-2(p+1)A+2B-a\phi} &=& 0, \label{EqR} \\
r H' - 1 - (k-1) {\rm e}^{2B-2C} &=& 0. \label{EqO}
\end{eqnarray}

Obviously the equations (\ref{EqPhi}), (\ref{EqA}) and
(\ref{EqC}) are similar in structure; moreover, the physical
conditions to be imposed on the solutions are also similar,
therefore the function $C$ and the dilaton $\phi$ have to be
related to the function $A$ as follows
\footnote{ In the case $k=q-1$ the equation (\ref{EqC}) does not
arise, while the relation between $C$ and $A$ in (\ref{SolPhi})
can be treated as a gauge fixing for the radial coordinate $r$.
In the smeared cases, $k<q-1$, the metric ansatz (\ref{metric})
specifies the gauge completely (this metric pattern is not
preserved under a radial coordinate redefinition). }
\begin{equation}\label{SolPhi}
\phi = \frac{a(D-2)}{q-1} A, \qquad C = -\frac{p}{q-1} A.
\end{equation}
Consequently, the equations of motion reduce to a second order
differential equation for $A$
\begin{equation}
A'' + A'^2 - A'B' + k r^{-1} A' - \frac{2(q-1)b^2}{D-2} {\rm
e}^{-\lambda A+2B} = 0, \label{EqA''}
\end{equation}
and two equations in terms of $B'$ and $B$
\begin{eqnarray}
B' &=& A' + (k-1) r^{-1} - (k-1) r^{-1} {\rm e}^{2B +
\frac{2p}{q-1}A}, \label{EqB'} \\
{\rm e}^{2B} &=& \frac{2k(k-1)(q-1) + 4k(D-2)r A' -
(D-2)(\lambda-2) r^2 A'^2}{(q-1) [ 2k(k-1) {\rm
e}^{(\lambda+\frac{2p}{q-1})A} + 4 b^2 r^2 ] } {\rm e}^{\lambda
A}, \label{EqE2B}
\end{eqnarray}
where the parameter $\lambda$ is defined as
\begin{equation}
\lambda = 2(p+1) + \frac{a^2(D-2)}{q-1}.
\end{equation}

In fact, the three equations (\ref{EqA''}), (\ref{EqB'}) and
(\ref{EqE2B}), are not independent. Using the Eq. (\ref{EqE2B}),
$B$ can be expressed in terms of $r, A$ and $A'$. Once we
substitute either (\ref{EqE2B}) into (\ref{EqB'}), or
(\ref{EqE2B}) and (\ref{EqB'}) into (\ref{EqA''}), we obtain the
following non-linear second order differential equation for the
function $A$:
\begin{eqnarray}
& (k-1)(D-2)  [ 2k(q-1)r A'' + 2k^2(q-1) A' + 4k(D-2)r A'^2 -
(D-2)(\lambda-2)r^2 A'^3 ] {\rm e}^{(\lambda+\frac{2p}{q-1})A}&
\nonumber\\
& + 2 (q-1) b^2 r \left[ 2(D-2)r^2 A'' - 2(2k-1)(D-2) r A' +
(D-2)(\lambda-2) r^2 A'^2 - 2k(k-1)(q-1) \right] = 0.&
\label{EqKey}
\end{eqnarray}
Ultimately then, the construction of a fluxbrane in a general
dilatonic theory is reduced to solving the master equation
(\ref{EqKey}); the functions $B,\,C$ and $\phi$ being then
straightforwardly calculated from the Eqs. (\ref{EqE2B}) and
(\ref{SolPhi}), while the form field F$_{[q]}$ is readily given by
(\ref{SolF}). In the Table \ref{table}, we list the values of
parameters for eleven-dimensional supergravity and types IIA and
IIB theories in ten dimensions. The corresponding fluxbranes are
not entirely independent, the discrete S-duality (\ref{duality})
relates them in pairs; these electric/magnetic pairs are indicated
in parentheses. The column for the F4-brane of IIB supergravity
has been left blank because the five-form field strength should be
self-dual, this is not ensured by our ansatz. Note that for all
entries in this table the following relation is satisfied
\begin{equation}\label{relation}
\lambda -2=\frac{4(D-2)}{q-1}.
\end{equation}

\begin{table}
\caption{\label{table} Parameters of fluxbranes}
\begin{tabular}{|c||c|c||c|c|c|c|c|c|c|c|c|} \hline
  & \multicolumn{2}{|c|}{M-theory}
  & \multicolumn{9}{|c|}{Type II string theories} \\
  \hline
  \qquad & \quad F6 \quad & \quad F3 \quad & \quad F7 \quad &
  \quad NS F6 \quad & \quad F6 \quad & \quad F5 \quad & \quad F4 \quad &
  \quad F3 ($\ast$F5) \quad & \quad F2 ($\ast$F6)\quad &
  \quad NS F2 ($\ast$NS F6)\quad & \quad F1 ($\ast$F7)\quad \\
  \hline\hline
  D & 11 & 11 & 10 & 10 & 10 & 10 & 10 & 10 & 10 & 10 &10 \\
  \hline
  $q$ & 4 & 7 & 2 & 3 & 3 & 4 & 5 & 6 & 7 & 7 & 8 \\
  \hline
  $a$ & 0 & 0 & 3/2 & -1 & 1 & 1/2 &  & -1/2 &
  -1 & 1 &-3/2 \\
  \hline\hline
  $p$ & 6 & 3 & 7 & 6 & 6 & 5 &  & 3 & 2 & 2 &1 \\
  \hline
  $\lambda$ & 14 & 8 & 34 & 18 & 18 & 38/3 &  &
  42/5 & 22/3 & 22/3 & 46/7 \\
  \hline
\end{tabular}
\end{table}

We were unable to find globally regular solutions to the above
equation analytically, except for $k=1$ case, which is the
generalized Melvin solution. There exists, however, an analytic
solution singular at the origin which correctly reproduces the
behaviour of regular solutions in the asymptotic limit. For $k=1$
the global solution can be obtained in a closed form.

\section{Generalized Melvin}
One solvable case is the $k=1$ (maximally smeared) solution,
which corresponds to a generalization of the Melvin flux tube. In
this section, we give the general solution for this particular
case. In the context of eleven-dimensional supergravity similar
solutions have been discussed in \cite{ChGaSh99b}. The equation
for $A$ becomes simply
\begin{equation}
2 r^2 A'' - 2 r A' + (\lambda - 2) r^2 A'^2 = 0,
\end{equation}
its exact solution reads
\begin{equation}
A = \frac2{\lambda-2} \left[ \ln( 1 + c_1^2 r^2 ) + \ln c_2
\right],
\end{equation}
where $c_1$ and $c_2$ are arbitrary constants. For the metric
function $B$ one finds generally
\begin{equation}
B = A + \ln\left( \frac{c_1 c_2}{b} \right) + \frac12 \ln \left[
\frac{4(D-2)}{(q-1)(\lambda-2)} \right].
\end{equation}
In view of the relation (\ref{relation}) the argument of the
constant under the second term logarithm is equal to one,
therefore for all parameters listed in Table \ref{table} one has
$B=A+\ln(c_1c_2/b)$. The form field strength reads
\begin{equation}
F_{[q]} = 2 c_1 c_2 \sqrt{\frac{4(D-2)}{(q-1)(\lambda-2)}} \,\, r
\, {\rm e}^{-(\lambda-2) A} \, d r \wedge d\phi \wedge dy_1
\wedge \cdots \wedge dy_{q-2},
\end{equation}
and again the relation (\ref{relation}) implies that the square
root coefficient in the expression for the form field is equal to
one for the tabulated list of parameters. Without loss
generality, one can specify the values $c_1=b$ and $c_2=1$ which
recover the earlier known generalizations of Melvin flux tube.

One reason why the case $k=1$ is integrable apparently lies in
the enhanced U-duality symmetry of this family of solutions, in
fact the above solution can be generated algebraically via some
transformation in solution space \cite{ChGaSh99b}.

\section{Attractor Solution}
For $k\neq 1$ we were unable to find a globally regular solution
to the master equation (\ref{EqKey}), however a simple solution
generalizing that of \cite{Sa01a,GuSt01} can be found which
represents the asymptotic of the regular solutions with different
degrees of smearing. Examining the Eq. (\ref{EqKey}) one can
observe the scaling symmetry $r \to \Gamma r$, ${\rm e}^A \to
\Gamma^\alpha {\rm e}^A$ for some $\alpha$. In view of this one
can try the solution of the following form
\begin{equation}
\label{attract}
A = \alpha \ln r + \ln \beta,
\end{equation}
with constants $\alpha$ and $\beta$. This particular solution was
presented in \cite{Sa01a,GuSt01} for $k=q-1$, and, whilst by
itself it represents a singular space-time as $r\to 0$, it is the
attractor solution for fluxbranes which are regular at their
core. For $k\ne1$, the value of $\alpha$ can be simply determined
from (\ref{EqKey}) by comparing powers of $r$, this gives
\begin{equation}
\alpha = \frac{2(q-1)}{2p+(q-1)\lambda}.
\end{equation}
The corresponding value of $\beta$ is also easily determined as
\begin{equation}
\beta = \left[ \frac{2(q-1)b^2}{\alpha(k-1)(D-2)}
\right]^{\alpha/2}.
\end{equation}
Somewhat unexpectedly, the value of $\alpha$ is independent of
the parameter $k$, that is the growth of the metric at infinity
is the same for all smeared solutions $1<k\leq q-1$. From the Eq.
(\ref{EqE2B}) we obtain the following expression for $B$
\begin{equation}
B = -\frac{p}{q-1}\, A + \frac12 \ln \gamma,
\end{equation}
where
\begin{equation}
\gamma = \frac{2k(k-1)(q-1) + 4 k \alpha (D-2) - \alpha^2 (D-2)
(\lambda-2)}{2(k-1)[ k(q-1) + \alpha (D-2) ]}.
\end{equation}

It is worth noting that the asymptotic form of the generalized
Melvin solution $k=1$ is different from what we have found here.
Using the Eq.(\ref{relation}) to express parameters in terms of
the space-time dimension and the rank of the antisymmetric field
strength, one finds for $k\neq 1$
\begin{equation}
{\rm e}^A \sim r^{(q-1)/[3(D-2)]},
\end{equation}
while for $k=1$ one has
\begin{equation}
{\rm e}^A \sim r^{(q-1)/(D-2)}.
\end{equation}

\section{M- and String Theory fluxbranes }
In eleven-dimensional supergravity one has a three-form potential
with the standard kinetic term, but also a Chern-Simons term. The
latter, however, plays no role for the purely electric or purely
magnetic solutions we are dealing with, so our general framework
remains valid for describing these solutions. The four-form field
strength can support two different types of fluxbranes: the
magnetic F6-brane ($q=4$, $p=6$), and the electric F3-brane, the
S-dual of the F6-brane. The latter lies within our prescription
when working with the seven-form dual to the four-form, $q=7$,
$p=3$, see (\ref{duality}).

\subsection{F6 solutions}

The F6-branes correspond to parameters $D=11, a=0, q=4$ and $p=6$
in the Table \ref{table} with $k$ giving the level of smearing.
The master equation for $A$ (\ref{EqKey}) reads
\begin{equation}\label{F6eqn}
3 (k-1) {\rm e}^{18A} [ k r A'' + k^2 A' + 6 k r A'^2 - 18 r^2
A'^3 ] + 2 b^2 r [ 3 r^2 A'' - 3 (2k-1) r A' + 18 r^2 A'^2 -
k(k-1) ] = 0.
\end{equation}
The other variables can be calculated as follows
\begin{eqnarray}
{\rm e}^{2B} &=& \frac{k(k-1) + 6 k r A' - 18 r^2 A'^2}{ k(k-1)
{\rm e}^{18A} + 2 b^2 r^2} {\rm e}^{14A}, \\
C &=& - 2 A, \\
F_{[4]} &=& 2 b \, r^k \, {\rm e}^{-13A+B} \, d r \wedge
\epsilon_{[k]} \wedge dy_1 \wedge \cdots \wedge dy_{3-k}.
\end{eqnarray}

The attractor solution ($k\ne 1$) reads
\begin{eqnarray}
A &=& \frac19 \ln r + \frac1{18} \ln \left( \frac{6 b^2}{k-1}
\right), \\
B &=& -\frac29 \ln r - \frac19 \left( \frac{6 b^2}{k-1} \right)
+ \frac12  \ln \left( \frac{3k-2}{3k-3} \right), \\
C &=& - \frac29 \ln r - \frac19 \ln \left( \frac{6 b^2}{k-1}
\right).
\end{eqnarray}

The local series solution near the origin, for regular solutions,
starts as follows
\begin{equation}
A = A_0 \; + \; O(r^2),
\end{equation}
it contains (the only) free parameter $A_0$. We solved
(\ref{F6eqn}) using a fifth order Runge-Kutta technique, with
$A'(0)=0$ as the initial condition. Numerical solutions starting
with different $A_0$ converge to the same attractor solution in
the asymptotic region (Fig. \ref{fig:attractorF6}). From
(\ref{EqE2B}) and (\ref{SolPhi}) we see that $B(0)=C(0)$, showing
that there is no conical singularity at the core.

\begin{figure}
\includegraphics[width=8cm]{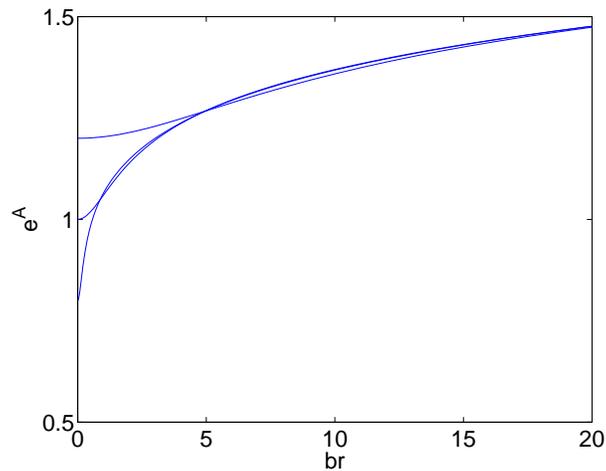}
\caption{\label{fig:attractorF6} Solutions to (\ref{F6eqn}) with
varied initial conditions, showing the attractor nature of the
solutions in the asymptotic region.}
\end{figure}

\subsection{F3 solutions}

The electric F3-branes correspond to $D=11, a=0, q=7$ and $p=3$
in the Table \ref{table}; the master equation for $A$ becomes
\begin{equation}\label{F3eqn}
3(k-1) {\rm e}^{9A} [ 2 k r A'' + 2 k^2 A' + 6 k
r A'^2 - 9 r^2 A'^3 ] + 4 b^2 r [ 3 r^2 A'' - 3 (2k-1) r A' + 9
r^2 A'^2 - 2 k(k-1) ] = 0,
\end{equation}
other variables in terms of $A$ read
\begin{eqnarray}
{\rm e}^{2B} &=& \frac{2k(k-1) + 6 k r A' - 9 r^2 A'^2}{ 2k(k-1)
{\rm e}^{9A} + 4 b^2 r^2} {\rm e}^{8A}, \\
C &=& - \frac12 A, \\
F_{[7]} &=& 2 b \, r^k \, {\rm e}^{-7A+B} \, d r \wedge
\epsilon_{[k]} \wedge dy_1 \wedge \cdots \wedge dy_{6-k}.
\end{eqnarray}
Similarly, the attractor solution for ($k\ne 1$) is
\begin{eqnarray}
A &=& \frac29 \ln r + \frac19 \ln \left( \frac{6 b^2}{k-1}
\right), \\
B &=& -\frac19 \ln r - \frac1{18} \left( \frac{6 b^2}{k-1} \right)
+ \frac12  \ln \left( \frac{3k-2}{3k-3} \right), \\
C &=& - \frac19 \ln r - \frac1{18} \ln \left( \frac{6 b^2}{k-1}
\right).
\end{eqnarray}

We solved this system numerically for the same initial conditions
$A_0=1,\;A'(0)=0$, and the various $k$ ($k=2,...,6$) to observe
the effect of smearing. The $k=1$ solution, given in Sec. III, is
known analytically and is shown as the dotted curve in Fig.
\ref{fig:attractorF3}. The lower curves show the behaviour of
solutions with $k=$ 2 (the top solid curve), 3, 4, 5, 6 (the
bottom curve).

\begin{figure}
\includegraphics[width=8cm]{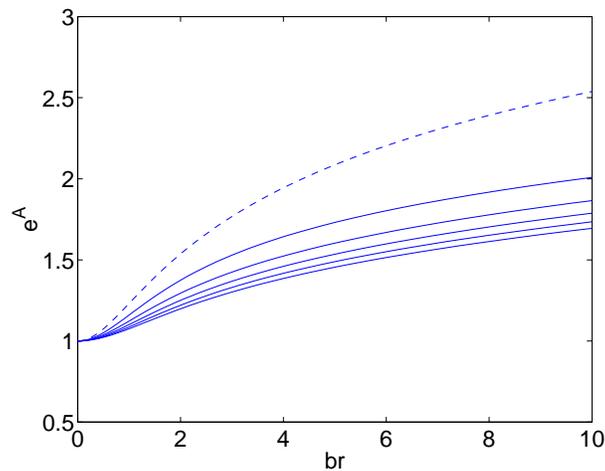}
\caption{\label{fig:attractorF3} Solutions to (\ref{F3eqn})
with different levels of smearing, the upper (dashed) curve is
for $k=1$ with the lower curves describing $k=2,3,4,5,6$.}
\end{figure}

\subsection{Type II fluxbranes}

The NS-NS sectors of type IIA and IIB theories are the same, with
the three-form field strength coupled to the dilaton with the
coupling constant $a=-1$. Therefore, there exist  magnetic NS
F6-branes with $k=1,\,2$. The dual, electric, NS F2-branes
 come  through the discrete duality symmetry of
(\ref{action}), for them $k=1,...,6$.

In the R-R sector there are different rank $q$-form fields: $q$ is
even for IIA and odd for IIB. The dilaton coupling constant in the
Einstein frame assumed here is $a=(5-q)/2$ and we see from
(\ref{duality}) that these fluxbranes come in dual pairs. There is
one exceptional case, not covered by our treatment: the five-form
field strength of type IIB supergravity is self dual and our
analysis does not cover this case of a dyon like solution. Table
\ref{table} shows a summary of the above discussion.

\section{Conclusion}
We have shown that apart from the already known supergravity
fluxbrane solutions with the spherically symmetric transverse
space there exist a sequence of fluxbranes with a cylindrical
transverse space being the product of a lower dimensional sphere
and a flat space. All such solutions are shown to be governed by a
single second order radial differential equation for the metric
function. This improves on the previous analyses where solutions
were described in terms of two unknown functions. An analytic
solution was presented which has the correct asymptotic behaviour
but diverges at the origin. Solutions which are regular at the
origin were shown to depend on one parameter and tend to the
analytic solution for sufficiently large radius. This gives
numerical evidence for existence of globally regular solutions for
all transverse topologies considered. The radial growth at
infinity for the solution is the same for all degrees of smearing
apart from the exceptional case of maximal smearing. For this an
analytic solution exists which can also be obtained algebraically
using U-duality.

\begin{acknowledgments}
We would like to thank Dominic Brecher, Ed Copeland, Bert Janssen
and Hyun Seok Yang for discussions. The work of CMC was supported
by the CosPA project, the National Science Council, Taiwan, and
in part by the Center of Theoretical Physics at NTU and the
National Center for Theoretical Sciences. The work of DVG was
supported by the Russian Foundation for Basic Research under
grant 00-02-16306. The work of PMS was funded by PPARC.
\end{acknowledgments}



\begin{thebibliography}{99}

\bibitem{Me64}
    M.A. Melvin,
    {\sl Pure magnetic and electric geons},
    {\it Phys. Lett. \bf 8} (1964) 65.

\bibitem{GiMa88}
    G.W. Gibbons and K. Maeda,
    {\sl Black holes and membranes in higher dimensional theories
         with dilaton fields},
    {\it Nucl. Phys. \bf B298} (1988) 741-775.

\bibitem{DoGaKaTr94}
    H.F. Dowker, J.P. Gauntlett, D.A. Kastor and J. Traschen,
    {\sl Pair creation of dilaton black holes},
    {\it Phys. Rev. \bf D49} (1994) 2909-2917;
    {\tt hep-th/9309075}.

\bibitem{DoGaGiHo95}
    H.F. Dowker, J.P. Gauntlett, G.W. Gibbons and G.T. Horowitz,
    {\sl Decay of magnetic fields in Kaluza-Klein theory},
    {\it Phys. Rev. \bf D52} (1995) 6929-6940;
    {\tt hep-th/9507143}.

\bibitem{DoGaGiHo96}
    H.F. Dowker, J.P. Gauntlett, G.W. Gibbons and G.T. Horowitz,
    {\sl Nucleation of $p$-branes and fundamental strings},
    {\it Phys. Rev. \bf D53} (1996) 7115-7128;
    {\tt hep-th/9512154}.

\bibitem{CoGu00}
    M.S. Costa and M. Gutperle,
    {\sl The Kaluza-Klein Melvin solution in M-theory},
    {\it JHEP \bf 0103} (2001) 027;
    {\tt hep-th/0012072}.

\bibitem{GaRy98}
    D.V. Gal'tsov and O.A. Rytchkov,
    {\sl Generating branes via sigma-models},
    {\it Phys. Rev. \bf D58} (1998) 122001;
    {\tt hep-th/9801160}.

\bibitem{ChGaSh99b}
    C.-M. Chen, D.V. Gal'tsov and S.A. Sharakin,
    {\sl Intersecting $M$-fluxbranes},
    {\it Grav. Cosmol. \bf 5} (1999) 45-48;
    {\tt hep-th/9908132}.

\bibitem{ChGaSh99a}
    C.-M. Chen, D.V. Gal'tsov and S.A. Sharakin,
    {\sl Vacuum interpretation for supergravity $M$-branes},
    {\it Phys. Lett. \bf B475} (2000) 269-274;
    {\tt hep-th/9908133}.

\bibitem{Sa01a}
    P.M. Saffin,
    {\sl Gravitating fluxbranes},
    {\it Phys. Rev. \bf D64} (2001) 024014;
    {\tt gr-qc/0104014}.

\bibitem{GuSt01}
    M. Gutperle and A. Strominger,
    {\sl Fluxbranes in string theory},
    {\it JHEP \bf 0106} (2001) 035;
    {\tt hep-th/0104136}.

\bibitem{CoHeCo01}
    M.S. Costa, C.A.R. Herdeiro and L. Cornalba,
    {\sl Flux-branes and the dielectric effect in string theory},
%
    {\tt hep-th/0105023}.

\bibitem{Em01}
    R. Emparan,
    {\sl Tubular branes in fluxbranes},
    {\it Nucl. Phys. \bf B 610} (2001) 169-189;
    {\tt hep-th/0105062}.

\bibitem{Sa01b}
    P.M. Saffin,
    {\sl Fluxbranes from $p$-branes},
%
    {\tt hep-th/0105220}.

\bibitem{brecher01}
    D. Brecher and P.M. Saffin,
    {\sl A note on the supergravity description of dielectric branes.},
    {\it Nucl. Phys. \bf B 613} (2001) 218-236;
    {\tt hep-th/0106206}.

\bibitem{Mo01}
    L. Motl,
    {\sl Melvin matrix models},
%
    {\tt hep-th/0107002}.

\bibitem{Su01}
    T. Suyama,
    {\sl Melvin background in hetrotic theories},
%
    {\tt hep-th/0107116}.

\bibitem{Ur01}
    A.M. Uranga,
    {\sl Wrapped fluxbranes},
%
    {\tt hep-th/0108196}.
\bibitem{RuTs01}
    J.G. Russo and A.A. Tseytlin,
    {\sl Magnetic backgrounds and tachyonic instabilities in
         closed superstring theory and M-theory},
    {\it Nucl. Phys. \bf B 611} (2001) 93-124;
    {\tt hep-th/0104238}.

\bibitem{Ts01}
    A.A. Tseytlin,
    {\sl Magnetic backgrounds and tachyonic instabilities in
         closed string theory},
%
    {\tt hep-th/0108140}.

\end{thebibliography}
\end{document}